# Pervasive liquid metal direct writing electronics with roller-ball pen

Yi Zheng [1], Qin Zhang [1], and Jing Liu[1, 2, a)]

[1] Beijing Key Lab of CryoBiomedical Eng. and Key Lab of Cryogenics,
Technical Institute of Physics and Chemistry,
Chinese Academy of Sciences, Beijing 100190, China

[2] Department of Biomedical Engineering, School of Medicine, Tsinghua University,
Beijing 100084, China

A roller-ball pen enabled direct writing electronics via room temperature liquid metal ink was proposed. With the rolling to print mechanism, the metallic inks were smoothly written on flexible polymer substrate to form conductive tracks and electronic devices. The contact angle analyzer and scanning electron microscope were implemented to probe the inner property of the obtained electronics. An ever high writing resolution with line width and thickness as 200μm and 80μm, respectively was realized. Further, with the administration of external writing pressure, $GaIn_{24.5}$ droplets embody increasing wettability on polymer which demonstrates the pervasive adaptability of the roller-ball pen electronics.

---
[a)] Author to whom correspondence should be addressed.
Electronic mail: jliu@mail.ipc.ac.cn; Tel.: 86-10-82543765.

Over the past decade, flexible printed electronics have been increasingly extended to the fabrication of transistors [1], displays [2], sensors [3, 4], antennas [5, 6], and photovoltaics [7] etc. Along with developing the ink materials, diverse printing technologies such as micro-contact printing, nano-imprinting, screen printing, roll-to-roll printing and inkjet printing etc. were also proposed to make desired device. Among them, the cost effective and convenient approaches to rapidly manufacture flexible electronics were specially paid with great attentions. Recently, an alternative strategy to directly write out conductive tracks on Xerox paper substrate was reported using a roller-ball pen filled with conductive silver ink. [8] The scientists there demonstrated that such writing was capable of offering a distinct way to draw flexible devices, which contributes significantly to the emerging area of the direct electronics fabrication. However, the inherent limitation of the technology also restricts it from being pervasively used. For example, it is still somewhat cumbersome and complicated to configure the silver ink with satisfactory attributes for the well-function of roller-ball pen. Most importantly, the acquired electrical performance of the silver ink (already loaded with up to 50 wt% silver particles) is still not conductive enough, especially when high temperature post-treatmentwas not administrated. Besides, the resolution of the written conductive tracks (approximately 650 μm) still needs further improvement. Clearly, finding a better pen electronics with significantly improved conductivity (preferably metal-like property) in room temperature and appropriate fluidity for higher writing resolution is urgently required to fulfill the increasing need of direct writing electronics.

Recently, as a new class of functional materials in printed electronics area, the room temperature liquid metals were proposed [9, 10] as printing inks owing to their appealing writable properties, favorable metallic conductivity, moderate cost and environment-friendly property. Along this way, several matched direct-printing technologies were developed. [11, 12] However, until now, there seriously lacks of a highly convenient and portable tool for the fluent writing of such promising inks, just as an ordinary pen does. The mechanism can be attributed to the difficulty that the large surface tension of the metal inks raised big challenges for its high quality delivery and printing.

Here, through resolving the fundamental issues lying behind, we demonstrate that it is feasible to directly and instantly write out various conductive structures on flexible polymer substrate with the room temperature metallic inks. A rolling to print mechanism was clarified to be highly compatible for writing the metal inks and thus forming various desired electronic devices on flexible substrate. This may lead to the pervasive electronics in the sense of that writing conductive line looks just as easy as signing a name or drawing a picture on the paper. For demonstrating the working principle, the binary eutectic alloy of gallium and indium was adopted as the printed ink. It is natural to accept that the roller-ball pen owns the dominant merits as wide popularity, favorable portability and low-cost. In addition, the easily available pens with varied ball diameters ranging from 200 μm to 1000μm are rather beneficial for carrying out necessary comparative experiments. This would help develop pervasive pens electronics for future daily use through justifying the writing precision. In this letter, a liquid metal roller-ball pen (LMRP) was developed and applied to write conductive typical tracks and electrical devices on flexible polymer substrates. Post-analysis was carried out to interpret such writing mechanisms, as well as the availability and performance of the written circuits and devices.

For configuring the room temperature liquid metal ink, here high-purity gallium and indium (with purity of 99.99 percent) metals as source materials were weighted with a ratio of 75.5:24.5 in line to make $GaIn_{24.5}$ alloy. The basic preparation processes were as follows: The weighted gallium and indium metals according to the required percentage are mixed in the beaker which was treated by deionized water in advance and heated up to 50 ℃ until metals were fused completely, then stirred slightly just for 30 seconds, thus the configuration process is completed. After that, the well-prepared $GaIn_{24.5}$ ink was injected into the empty tube of the roller-ball pen to serve as the liquid metal ink.

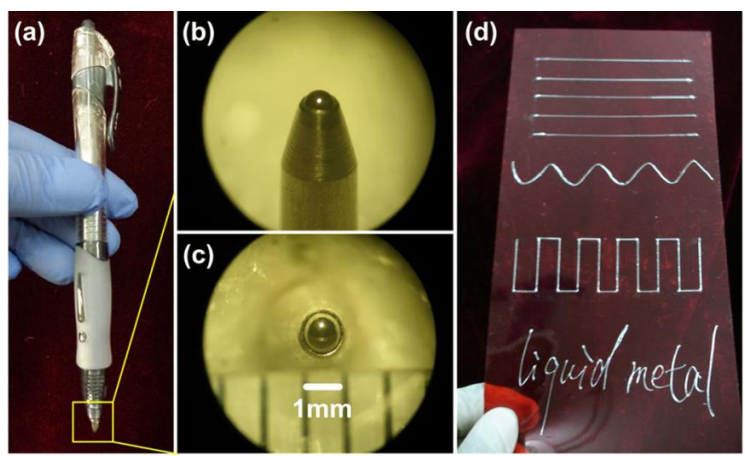

FIG.1. (a) Optical image of a roller-ball pen loaded with liquid metal ink. (b) and (c) Optical microscope images of the side and top views of the pen nib. (d) Image of the conductive tracks written by the LMRP, containing lines, curves, fold line and words.

With the well-manufactured LMRP, a group of typical conductive tracks can be easily written on the flexible polymer substrate which is just as simple as drawing a picture on paper. Fig. 1(a) gives the optical image of a representative roller-ball pen with a ball diameter of approximately 950 μm that is filled with $GaIn_{24.5}$ ink. Fig. 1(b), (c) exhibit the optical microscope images of the nib, from which the structure can be clearly observed. Fig. 1(d) presents the conductive tracks smoothly written on the flexible polymer, containing conductive lines, curves, fold line and metal words ("liquid metal"), implying that the $GaIn_{24.5}$ ink owns commendable compatibility with the flexible polymer substrate. In order to thoroughly disclose the micro-morphology of the track, scanning electron microscope evaluation was also carried out. Fig. 2 reveals the SEM images of the liquid metal tracks on the polymer. It can be obviously seen from Fig. 2(a) that the metal track owns excellent uniformity on the polymer with a width of as tiny as 200 μm. Fig. 2(b) illustrates the SEM image regarding thickness of the $GaIn_{24.5}$ track from the side view, which reaches the level of 80μm. All these fine structures disclosed the highly acceptable writing precision of the liquid metal based roller-ball pen electronics, which is an important progress over existing electronics writing resolution, say about 600μm. Clearly, through innovating a finer size and structure of the roller-ball pen, a higher electronics writing resolution can still be possible in the coming future.

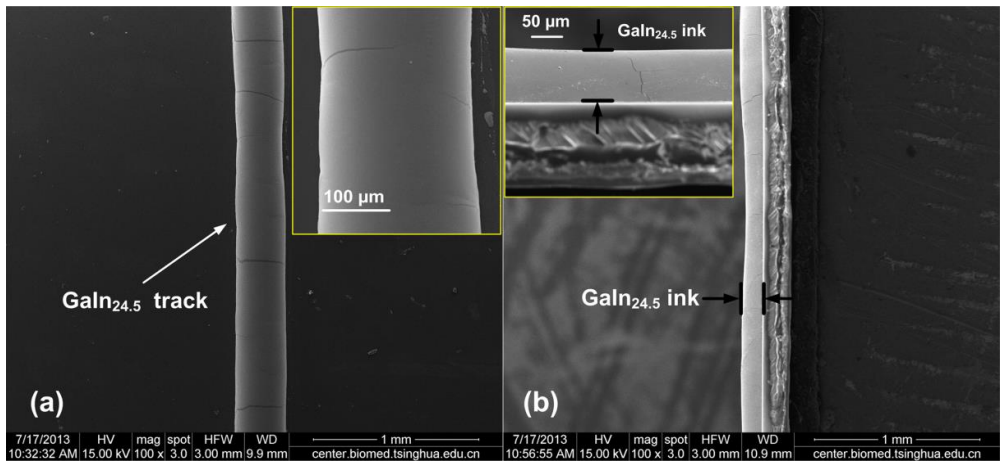

FIG.2. (a) SEM image of width of the $GaIn_{24.5}$ track written on the polymer substrate from the top view (inset reveals a larger magnification times of the track). (b) SEM image of thickness of the $GaIn_{24.5}$ track from the side view (inset also shows a larger magnification times).

In our research, the compatibility between liquid metal and substrate is found to be the key factor for the high quality electronics writing. It has been widely accepted that the contact angle $\Theta$ can be used to quantify the wettability of the liquids on substrates. Along this way, a contact angle meter (JC2000D3, Shanghai) was employed to measure the contact angle of $GaIn_{24.5}$ on polymer substrate under different external pressures. It can be obviously seen from Fig. 3 (a) that the contact angle of $GaIn_{24.5}$ on polymer (acquired through five points fitting method) decreases with the increasing of the tiny external pressures generated from the Polytetrafluoroethylene (PTFE) weighted 0.5, 1.0, 1.5, 2.0g, respectively. Meanwhile, according to the Young's equation, there exists a widely accepted physical discipline between the wettability and contact angle: $\Theta < 90$, indicating the liquid is capable of wetting the surface; $\Theta > 90$, indicating the liquid is unable to wet the surface. And, the smaller the contact angle $\Theta$, the better the wettability. Apparently, just given a tiny external pressure, the $GaIn_{24.5}$ droplets would favorably wet the flexible polymer substrate. This guarantees the present method a pervasive electronics writing way which can be used by an ordinary user without particular trained experience.

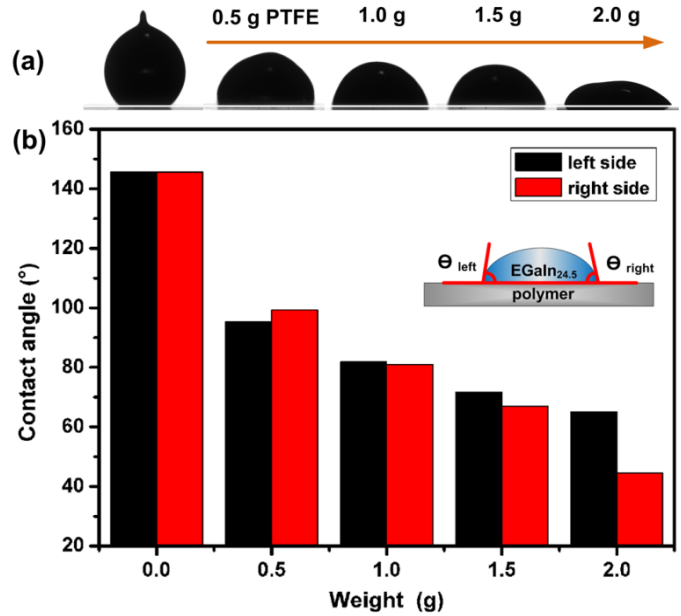

FIG. 3. (a) Optical images of GaIn24.5 droplet under different external pressure generated from the Polytetrafluoroethylene (PTFE) weighted 0.5, 1.0, 1.5, 2.0g, respectively. (b) Acquired left and right contact angle of $GaIn_{24.5}$ on polymer substrate using five points fitting method, illustrated image shows the schematic of left and right contact angle.

Originally, pure $GaIn_{24.5}$ displays poor wettability on the polymer due to its large surface tension. [13] Upon exposed to air, the $GaIn_{24.5}$ would quickly form a thin layer of gallium oxide on its surface, which guarantees the ink's good wetting with the substrate. Meanwhile, the whole electrical resistance of the inks remains largely unaffected since the thin layer owns the ability to attach the droplets and generates a high adhesion which contributes significantly to the compatibility on substrates. A tiny external force can easily deform the droplets and evidently improve the compatibility of $GaIn_{24.5}$ on polymer. When written, under the action of its own larger gravity (own a density as high as 6280 kg/m$^3$) and the rolling motion of the roller-ball, the $GaIn_{24.5}$ liquid ink with low-viscosity (around $2.7 \times 10^{-7}$ m$^2$/s) [14] can smoothly flow through the nib. Meanwhile the roller-ball offers a much larger pressure (than the pressure generated from the PTFE weighted 2g) on the effluent liquid metal ink, thus resulting in an excellent compatibility and subsequently a smoothly writing process.

It is particularly noteworthy that $GaIn_{24.5}$ owns a highly acceptable metallic electrical resistivity of approximately $2.98 \times 10^{-7}$ Ω·m [14] (Two orders of magnitude lower than silver inks at room temperature) [8], indicating its significant potential values in the coming printed electronics. With the help of various patterning instruments, various designed conductive structures or devices can be written out on the flexible polymer. Fig. 4 exhibits the images of several typical written metal structures through the LMRP. Fig. 4(a) exhibits the drawn conductive tracks written on polymer by another roller-ball pen with a roller-ball diameter of 700μm under help of a straightedge. Fig. 4(b) shows a parallel circuit of LED. Fig. 4(c) and (d) depict a written wire array and its excellent flexibility. Fig. 4(e) presents two directly–drawn capacitors with different sizes and measured values (through a digital electric bridge): $C_1$=2.0 pF, $C_2$=0.55 pF, respectively. Certainly, those structures can be succinctly and soundly packaged through printing the Polydimethylsiloxane (PDMS) to cover the objects if needed.

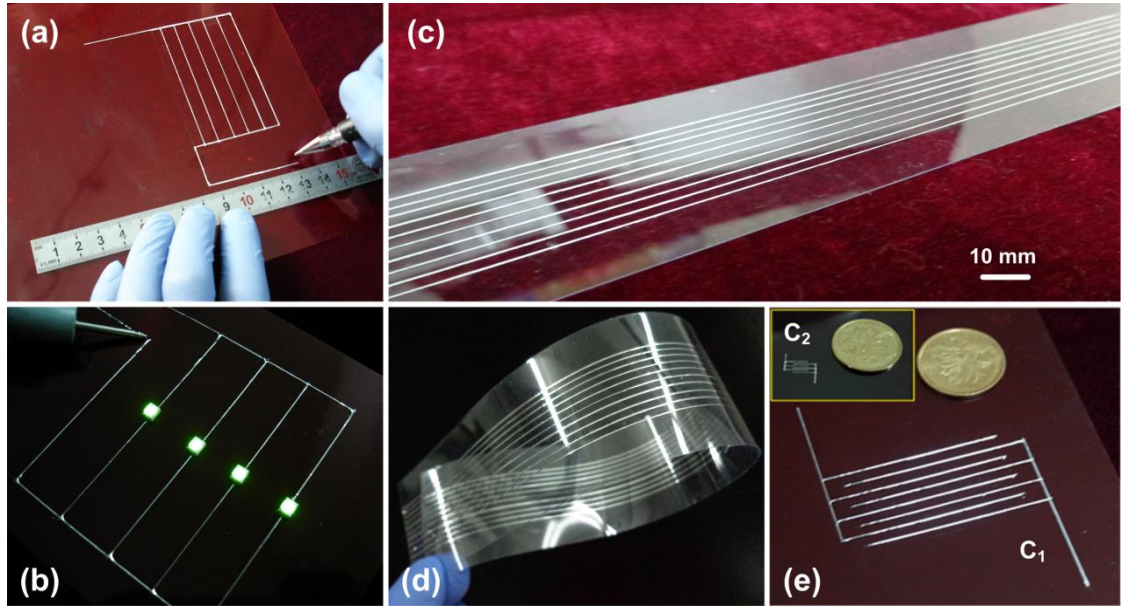

FIG. 4 (a) Drawn conductive tracks written on polymer by a roller-ball pen with a roller-ball diameter of 700μm with the help of a straightedge. (b) A simple parallel circuit of LED. (c) A written wire arrays. (d) The excellent flexibility of the written wire arrays. (e) Two direct–draw capacitors with different sizes and measured values: $C_1$=2.0 pF, $C_2$=0.55 pF.

In summary, with features of favorable writing resolution, attractive portability and metallic conductivity, the developed liquid metal roller-ball pen electronics owns promising potential for future printed electronics, especially under situations where rapid manufacturing, consumer oriented electronics, personalized electrical designing and DIY (Do It Yourself) electronics are urgently requested. Furthermore, through additional efforts, the liquid metal ink pen can still be configured to be loaded with multiple inks composed of diverse liquid metal and nano particles with prescribed physical or chemical properties. This will enhance the capability of the roller-ball pen electronics. In the near future, combined with the designed apparatus as well as matched control software, more electronic devices especially high precision electronic components such as complicated printed circuit board, miniature antennas, even batteries, etc. can be rapidly fabricated on flexible polymer.